# Single-Shot Quantitative X-ray Imaging Using a Primary Modulator and Dual-Layer Detector

Linxi Shi[1*], N. Robert Bennett[1], Alexander Vezeridis[1], Nishita Kothary[1], Adam S. Wang[1]

[1]Department of Radiology, Stanford University, Stanford, CA 94305, USA

## Abstract

**Purpose**: Conventional x-ray imaging and fluoroscopy have limitations in quantitation due to several challenges, including scatter, beam hardening, and overlapping tissues. In this work, we propose single-shot quantitative imaging (SSQI) by combining the use of a primary modulator (PM) and dual-layer (DL) detector, which enables motion-free dual-energy (DE) imaging with scatter correction in a single shot.

**Methods**: The key components of our SSQI setup include a PM and DL detector, where the former enables scatter correction for the latter while the latter enables beam hardening correction for the former. The SSQI algorithm allows simultaneous recovery of two material-specific images and two scatter images using four sub-measurements from the PM encoding. The concept was first demonstrated using simulation of chest x-ray imaging for a COVID patient. For validation, we set up SSQI on our tabletop system and imaged acrylic and copper slabs with known thicknesses (acrylic: 0-22.5 cm; copper: 0-0.9 mm), estimated scatter with our SSQI algorithm, and compared the material decomposition (MD) for different combinations of the two materials with ground truth. Second, we imaged an anthropomorphic chest phantom containing contrast in the coronary arteries and compared the MD with and without SSQI. Lastly, to evaluate SSQI in dynamic applications, we constructed a flow phantom that enabled dynamic imaging of iodine contrast.

**Results**: Our simulation study demonstrated that SSQI led to accurate scatter correction and MD, particularly for smaller focal blur and finer PM pitch. In the validation study, we found that the root mean squared error (RMSE) of SSQI estimation was 0.13 cm for acrylic and 0.04 mm for copper. For the anthropomorphic phantom, direct MD resulted in incorrect interpretation of contrast and soft tissue, while SSQI successfully distinguished them and reduced RMSE in material-specific images by 38% to 92%. For the flow phantom, SSQI was able to perform accurate dynamic quantitative imaging, separating contrast from the background.

**Conclusions**: We demonstrated the potential of SSQI for robust quantitative x-ray imaging. The simplicity of SSQI may enable its widespread adoption, including radiography and dynamic imaging such as real-time image guidance and cone-beam CT.

## 1. Introduction

X-ray imaging is a widely used tool in diagnostic imaging and image-guided procedures. Conventional x-ray imaging struggles with well-known physical challenges, including scatter, beam hardening, and overlapping tissues, limiting its capability to perform quantitative imaging [1,2].

Among all the non-idealities, x-ray scatter is in particular a challenge, which adds undesired signal that biases the accuracy of quantitative analysis. Without mitigation or correction strategies, the scatter signals can easily exceed primary x-ray signals. As a result, scatter correction has been an active research area for decades, with numerous scatter correction methods proposed, including both hardware- and software-based methods. General methods include scatter rejection using anti-scatter grids [3] and scatter estimation through beam blocker measurements [4,5], analytical modeling [6–8], Monte Carlo simulation [9–12], artificial intelligence [13], or with a primary modulator (PM) [14]. In particular, the most attractive feature about the primary modulator is its ability to simultaneously capture both primary and scatter images. The pre-patient PM uses alternating semitransparent and transparent regions to encode the primary image into a known checkerboard pattern, while the scatter image is largely unaffected due to its low-frequency nature. While a promising solution, one of the main challenges toward its adoption has been the beam hardening effect caused by the semitransparent regions of the PM, resulting in incomplete removal of the checkerboard pattern. Dual energy (DE) imaging, with its capability to characterize the object and PM attenuation across the entire spectrum and quantify area density of specific materials, is well-suited to address this issue, but only if the dual-energy projections are acquired with perfect spatial and temporal alignment. In addition to addressing the PM beam hardening, DE imaging has clinical benefits in bone removal, soft tissue removal, iodine-specific imaging for maskless digital-subtraction angiography (DSA), metal detection, electron density, effective atomic number, and virtual monoenergetic imaging [1,2][15–18]

DE imaging can be implemented in several ways, such as fast kV switching [19][20], fast filter switching [21], or a dual-layer (DL) detector [22]. While fast kV switching and fast filter switching have better spectral separation, they require a generator or complex hardware design and can still suffer from spatial and temporal misalignment between DE projections due to respiratory and cardiac motion, or due to system motion. On the other hand, a DL detector implicitly acquires DE projections with perfect spatial and temporal registration that completely removes the motion artifact, where the top layer tends to absorb low-energy photons while the high-energy photons are more likely to be absorbed by the bottom layer. However, there is no doubt scatter contamination remains a fundamental challenge for DL detectors and DE imaging in general, especially in large-area imaging that is routinely used in radiography, C-arm assisted image-guided procedures, or dedicated cone-beam CT (CBCT) systems.

Therefore, we propose to combine the use of a PM and DL detector to achieve single-shot quantitative imaging (SSQI). Previously, we have demonstrated a proof-of-concept using simulation studies [17][23]. In this work, we report a comprehensive investigation from simulation studies to realistic acquisitions of anthropomorphic and dynamic phantoms.

## 2. Methods and Materials

The proposed method combines the use of a PM and a DL flat-panel detector to achieve SSQI. The PM is used to perform scatter correction for the DL detector, where MD is sensitive to scatter signals, especially for radiography when a large field of view is acquired. At the same time, the DL detector can inherently correct for beam hardening in the PM by incorporating the known MD of the PM during single-shot material decomposition. Thus, the two combined technologies enable scatter-corrected dual-energy imaging in a single shot.

### 2.1 SSQI Setup

Figure 1 shows an overview of our SSQI setup. The PM has a checkerboard pattern with alternating transparent and semitransparent regions. The semitransparent region comprises of copper with 210 µm thickness, which reduces the intensity of a 120 kVp beam by approximately 50%. The original x-ray spectrum is modulated by the alternating regions, where the primary image is encoded with a checkerboard pattern whereas the low-frequency scatter image adds to it without any modulation. One of the main challenges toward its adoption has been the residual checkerboard pattern due to beam hardening [24]. We propose to solve this problem using motion-free dual-energy imaging enabled by the DL detector, which captures the combined modulated primary and unmodulated scatter images. In this work, the DL detector comprises a top CsI scintillator thickness of 200 µm coupled to an amorphous silicon flat-panel detector, a 1 mm Cu filter to increase spectral separation, and a bottom CsI scintillator thickness of 550 µm coupled to a second flat-panel detector. Both detectors have a native pixel size of 0.150×0.150 mm$^2$ and cover an active area of 43×43 cm$^2$. The top layer of the detector tends to absorb the low-energy (LE) photons and generates a LE projection, while the bottom layer absorbs more high-energy (HE) photons to generate a HE projection. Thus, the DL detector provides motion-free DE acquisition with perfect temporal alignment between the LE and HE projections. Spatial alignment is performed by an affine transformation to align and demagnify the bottom image to the top image, where the transformation is obtained via a one-time geometric calibration. A detailed description of this DL detector can be found in Ref [22].

The combined use of a PM and DL detector allows simultaneous estimation of scatter signals and material decomposition in a single exposure.

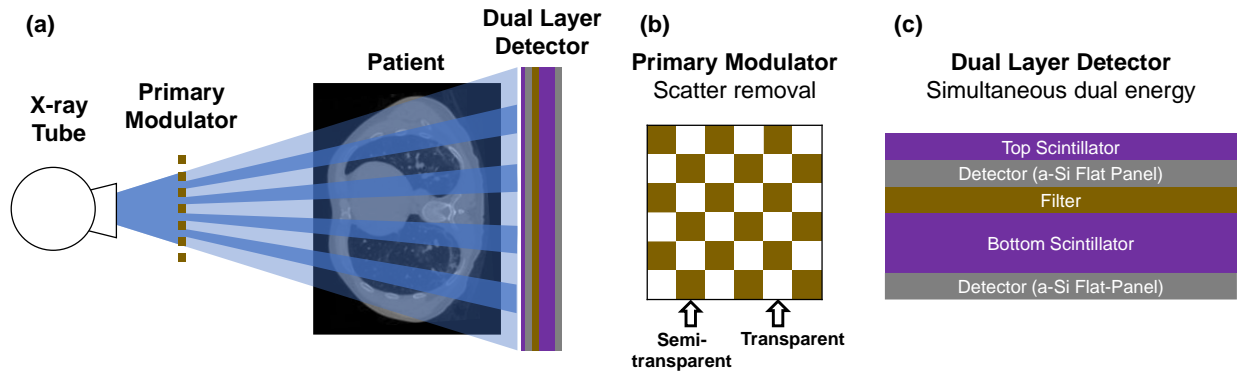

**Figure 1.** (a) The SSQI system comprises a primary modular and a dual layer detector. (b) The PM has a checkerboard pattern that enables scatter image estimation. (c) The DL detector generates dual energy images in a single shot.

## 2.2 SSQI Theory

The physical process of SSQI can be modeled as a polyenergetic source attenuated by the PM and object and detected by the DL detector. In this work, the object is assumed to be composed of two basis materials: acrylic and copper. The two basis materials were selected for their practicality in the calibration process and can be transformed into any other pairs of low/high Z materials. The SSQI algorithm pipeline fully utilizes the PM-encoded measurements from the two detector layers. Note that the proposed algorithm is based on two assumptions: 1) the scatter images have lower spatial frequency than the modulator pattern, and 2) the primary images of the object are approximately the same for neighboring semitransparent and transparent regions when downsampled and interpolated at low spatial resolution.

The DL images (top/bot) obtained behind the semitransparent and transparent regions of the PM (st/t) can be further divided into 4 sub-measurements (Figure 2), denoted as $M_{\text{top/bot, st/t}}$.

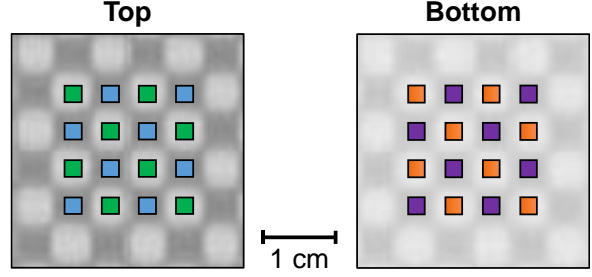

**Figure 2.** Magnified dual energy images of the primary modulator, where we can further derive additional measurements by dividing the signals measured behind the alternating semitransparent and transparent regions of the PM.

Mathematically, the 4 measurements can be expressed as a combination of corresponding primary, scatter, and noise images, as shown in Eqs. (1)-(4), where $I_{\text{eff,top/bot}}$ are the effective spectra corresponding to the top and bottom layers of the detector; $\mu_a(E)$ and $\mu_c(E)$ are the attenuation coefficients of two basis materials: acrylic and copper, and their corresponding line integrals for the object are $L_{a,\text{obj}}$, $L_{c,\text{obj}}$, respectively. The scatter images $S_{\text{top/bot}}$ are added to the modulated primary images and stay unchanged at each layer regardless of the modulation pattern due to its low-frequency nature. $N$ is the noise image for each measurement. Since the composition of the PM can be measured beforehand, the $L_{c,\text{PM}}$ is known *a priori*. Therefore, with four unknowns, $L_{a,\text{obj}}$, $L_{c,\text{obj}}$, $S_{\text{top}}$, and $S_{\text{bot}}$ in four measurements, the MD and scatter distributions can be easily estimated by solving Eqs. (1)-(4):

$$M_{\text{top,st}} = \int I_{\text{eff,top}}(E) e^{-\mu_a(E) L_{a,\text{obj}} - \mu_c(E)(L_{c,\text{obj}} + L_{c,\text{PM}})} dE + S_{\text{top}} + N \quad (1)$$

$$M_{\text{top,t}} = \int I_{\text{eff,top}}(E) e^{-\mu_a(E) L_{a,\text{obj}} - \mu_c(E) L_{c,\text{obj}}} dE + S_{\text{top}} + N \quad (2)$$

$$M_{\text{bot,st}} = \int I_{\text{eff,bot}}(E) e^{-\mu_a(E) L_{a,\text{obj}} - \mu_c(E)(L_{c,\text{obj}} + L_{c,\text{PM}})} dE + S_{\text{bot}} + N \quad (3)$$

$$M_{\text{bot,t}} = \int I_{\text{eff,bot}}(E) e^{-\mu_a(E) L_{a,\text{obj}} - \mu_c(E) L_{c,\text{obj}}} dE + S_{\text{bot}} + N; \quad (4)$$

### 2.2.1 SSQI Workflow

In reality, it is challenging to measure $I_{\text{eff,top/bot}}$ accurately. Instead of directly solving Eqs. (1)-(4), an alternative implementation is to perform a one-time empirical calibration to map the log-normalized, scatter-free measured DL signals (i.e., $-\log(M_{top/bot}/M_{top/bot,0})$, where $M_{top/bot,0}$ are the air images) to a series of pairs of acrylic and copper thicknesses that will be encountered in practice. Separate 5th-degree bivariate polynomials ($f_{a/c}$) were fit to the calibration data to map the log-normalized DL signals to each material.

We first focus on estimating the low-frequency scatter images, followed by MD at the native spatial resolution. After subtracting the scatter estimates, the object MD ($L_{a/c,obj}$) behind adjacent semitransparent and transparent regions of the PM should be similar, especially when the images are downsampled to lower spatial resolution. As a result, the problem can be simplified to estimating the scatter distribution that makes the object MD most consistent between semitransparent and transparent regions. Since the scatter images are low frequency and slowly vary across the detector, we can downsample the DL images substantially for faster computation. Figure 3 shows the workflow of the SSQI method and can be summarized as follows:

a) Start with original DL images obtained with 2×2 binning for a pixel size of 0.300×0.300 mm²; determine the semitransparent/transparent regions of the PM on the detector from an image of only the PM.
b) Expand the DL measurements to 4 subsets by separating the semitransparent/transparent regions, denoted as $M_{top/bot, st/t}$.
c) Fill the void regions on $M_{top/bot, st/t}$ using bilinear interpolation.
d) Downsample $M_{top/bot, st/t}$ with original size of 1440×1440 pixels to 40×40 pixels.
e) Feed the downsampled $M_{top/bot, st}$ and $M_{top/bot, t}$ into the MD polynomials. For each downsampled pixel, optimize the top/bottom scatter values to minimize the absolute difference of object MD ($L_{a/c,obj}$) from the semitransparent and transparent measurements. The non-linear optimization problem was solved by Matlab's 'fsolve' function, based on the trust-region-dogleg algorithm [25].
f) Obtain the downsampled $S_{top/bot}$ as an output of the optimization problem.

The resulting $S_{top/bot}$ includes both low-frequency scatter signals and estimation errors, stemming mostly from noise and the MD process in the SSQI pipeline. Further processing is used to refine the scatter distributions, as shown in Figure 4:

a) First resample the initial scatter estimates to their original size of 1440×1440 pixels. To reduce the impact of estimation errors, we identify the true scatter region, $\Omega_s$, as:

$$\Omega_s = \{(i,j) \mid |\nabla S(i,j)| < T_g, S(i,j) \geq 0\}, \tag{5}$$

where $(i,j)$ is the pixel index on the detector, $\nabla$ calculates the scatter gradient, $T_g$ is the gradient threshold, empirically set at 80% of the mean value of $|\nabla S|$, and scatter must have a non-negative value.

b) A previously described local filtration algorithm [26] is used to estimate the final scatter image. Basically, the algorithm performs a low-pass filtration to the signals within $\Omega_s$, while interpolating the void regions outside $\Omega_s$. A 2D Gaussian filter with standard deviation 20 pixels was used in this work.

The final $S_{\text{top/bot}}$ are then deducted from the original measurements, and the full-resolution, scatter-corrected DL images are fed into the MD polynomials to obtain the quantitative line integrals of acrylic and copper through the combination of the PM and object. The known PM MD is then subtracted from the combined MD to obtain the object MD. It typically takes ~8 sec to process one pair of images without parallel computing or other code optimization when implemented in Matlab 2020a (Mathworks, Natick, MA) on a standard laptop (2.5 GHz Intel Core i7 Apple MacBook).

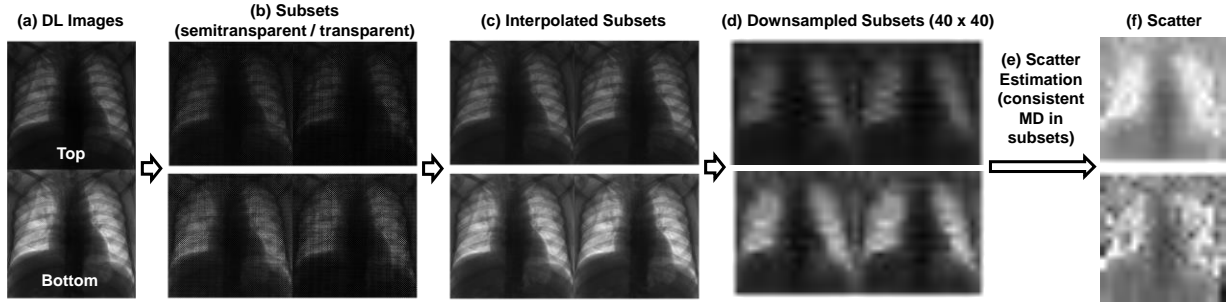

**Figure 3.** Workflow of scatter estimation. (a) DL images are (b) divided into subsets behind semitransparent / transparent regions of the PM, (c) interpolated in missing regions of each subset, and (d) downsampled. (e) Scatter is estimated such that the subsets produce consistent object MD. (f) Initial scatter estimates. (a)-(d) Window: [0 0.3], normalized to air; (f) window: [0 0.03].

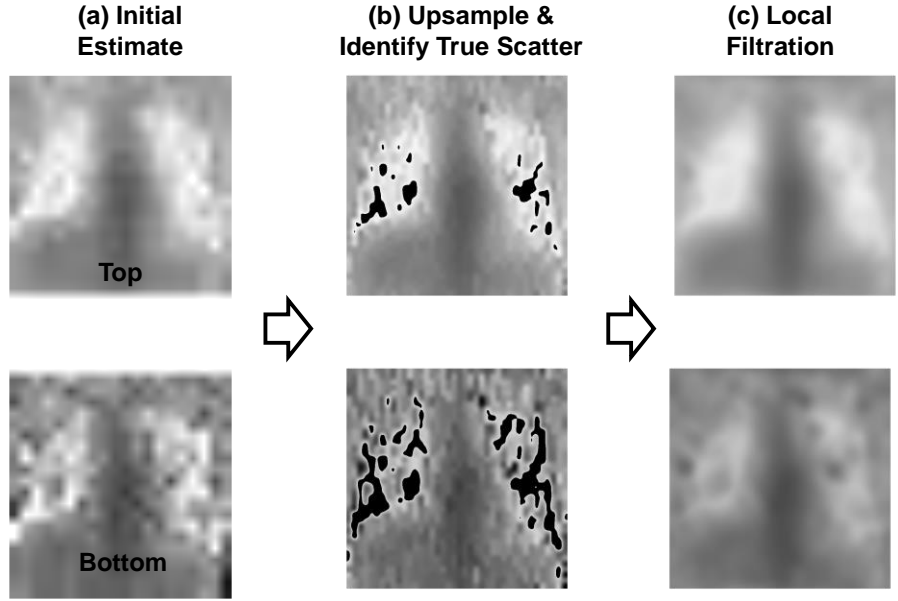

**Figure 4.** Refine the (a) initial scatter estimates by (b) upsampling and identifying true scatter regions, followed by (c) local filtration to smooth and fill in incorrect regions. Window: [0 0.03], normalized to air.

**2.3 Simulation**

To evaluate the feasibility of SSQI, we first performed a simulation study using chest CT images of a patient with COVID-19 infection to create a digital chest phantom [27], where the CT numbers were converted to mass density and classified into lung, soft tissue, and bone using a simple conversion curve with transition points at -200 and 200 HU between the materials. DL projections of the phantom were simulated using our SSQI setup, with a source-to-detector distance of 180 cm and the phantom placed in front of the detector, as is typical for chest x-ray. According to Eqs. (1)-(4), a polyenergetic 120 kVp, 5 mAs source was attenuated by the PM and the digital chest phantom, with additional consideration of the detector response and filtration modeled into effective spectra, $I_{\text{eff,top/bot}}$. Low-frequency scatter images were generated using the pep-model from each primary image [28], with a Gaussian filter with kernel size of 400×400 pixels and standard deviation of 200 pixels. The top layer scatter image was scaled assuming a maximum scatter-to-primary ratio (SPR) of 5, and the bottom layer SPR was reduced to 4.5 since scatter tends to be lower energy and less penetrating than primary. Poisson noise $N$ was added to the DL images to add realistic noise. Figure 5 shows the simulation pipeline to generate the DL measurements.

The PM images used for the simulation were acquired with realistic blur using an x-ray focal spot size of 0.3 mm on our tabletop system. The SSQI performance was directly related to the modulation capability of the PM, which was determined by a combined effect of focal blur and PM pitch size. In principle, finer pitch and smaller focal spot blur are preferred, however these options are limited by our system's hardware. To investigate the impact of these two factors, we acquired PM images with different pitch size (889 and 457 µm, based on previously manufactured PMs[14] and source-to-modulator distance (SMD: 23.1 and 36.2 cm, which are the minimum distance just outside the collimator and the maximum distance where the PM still covers the entire detector, respectively), where blur is reduced with larger SMD. The simulated DL projections were fed into our SSQI algorithm for evaluation.

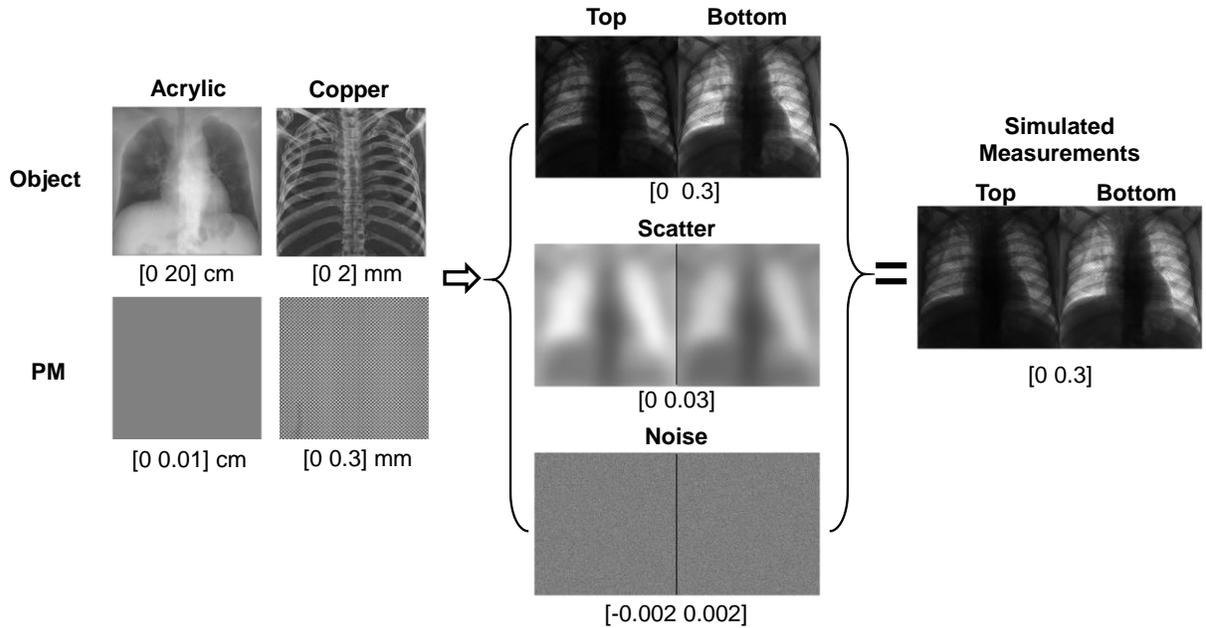

**Figure 5.** Simulation pipeline to generate the simulated DL images with scatter and noise. Middle and right images normalized to air.

## 2.4 Experiments
### 2.4.1 System Calibration and Validation

As mentioned previously, an empirical material decomposition method was used in SSQI: The DL detector measured attenuation from acrylic and copper in 42 pairs of thicknesses ranging from {0, 5.0, 10.0, 14.8, 19.9, 25.0} cm acrylic slabs and {0, 0.3, 0.6, 1.2, 1.8, 2.4, 3.0} mm copper sheets. The DE projections were acquired with an average of 64 exposures, each at 120 kV, 55 mA, and 20 ms on our tabletop system. The experimental setup is shown in Figure 6(a), where the DL detector and x-ray source were mounted on a tabletop system. To avoid scatter from the slabs, the source was collimated to 2×2 cm$^2$ at the detector and the calibration slabs were placed close to the source. A MD calibration matrix matching the measured DL attenuation and the actual material thicknesses were then acquired to fit MD polynomials for later use.

For validation, acrylic and copper thicknesses within the same range but at different combinations to the calibration dataset ({0, 5.9, 12.4, 17.5, 22.5} cm acrylic × {0, 0.3, 0.9} mm copper) were placed 30 cm away from the DL detector with the collimator open to 20×20 cm$^2$ at the detector to create realistic scatter signals, as shown in Figure 6(b). A PM with pitch 889 μm was placed at a SMD of 23.1 cm. Our SSQI algorithm was then used to estimate scatter and perform MD simultaneously. The resulting MD thicknesses were compared with the known material thicknesses for validation purposes.

### 2.4.2 Static SSQI

We further investigated the SSQI performance for dual-energy chest x-ray, using a static anthropomorphic chest phantom that contains barium contrast (similar to iodinated contrast) within the coronary arteries of a preserved heart. The phantom was placed 30 cm in front of the detector to mimic a chest x-ray or on an interventional room table (Figure 6(c)). To obtain reference images with minimal scatter, the beam was collimated to a small 2×2 cm$^2$ region on the detector, and the phantom was imaged without a PM.

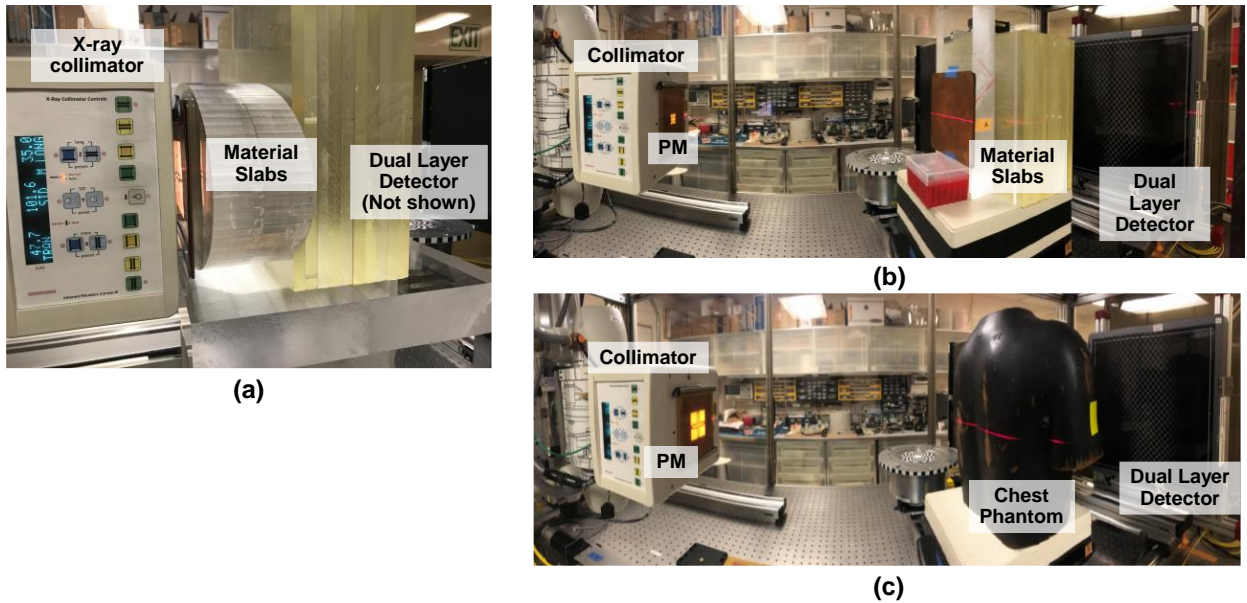

**Figure 6.** Experimental setup of SSQI for (a) material decomposition calibration, (b) material decomposition validation, and (c) dual-energy chest x-ray setup.

### 2.4.3 Dynamic SSQI

Another use case of SSQI is its potential for maskless digital subtraction angiography (DSA) in image-guided procedures. SSQI should be robust against motion artifacts since it does not require mask images. To test its performance in real time, we constructed a flow phantom that enabled dynamic imaging of iodine. The flow phantom consisted of a U-shape tube (inner diameter: 6.35 mm) encased in a uniform cylindrical solid-water phantom with a diameter of 7.5 cm. The phantom was pre-filled with water. An injection sequence of 50 ml of water at 15 ml/s (to clear any air bubbles in the system), followed by a contrast bolus of 50 ml of iodine at 10 ml/s with a concentration of 150 mg/ml (Omnipaque, GE Healthcare), followed by 100 ml of water at 10 ml/s to fully clear any contrast remaining in the flow phantom. Pulsed fluoroscopy (120 kV, 55 mA, 20 ms) and the DL detector were operated at 7.5 frames/sec (fps) to capture dynamic images of the flow. The SSQI setup for this experiment is shown in Figure 7.

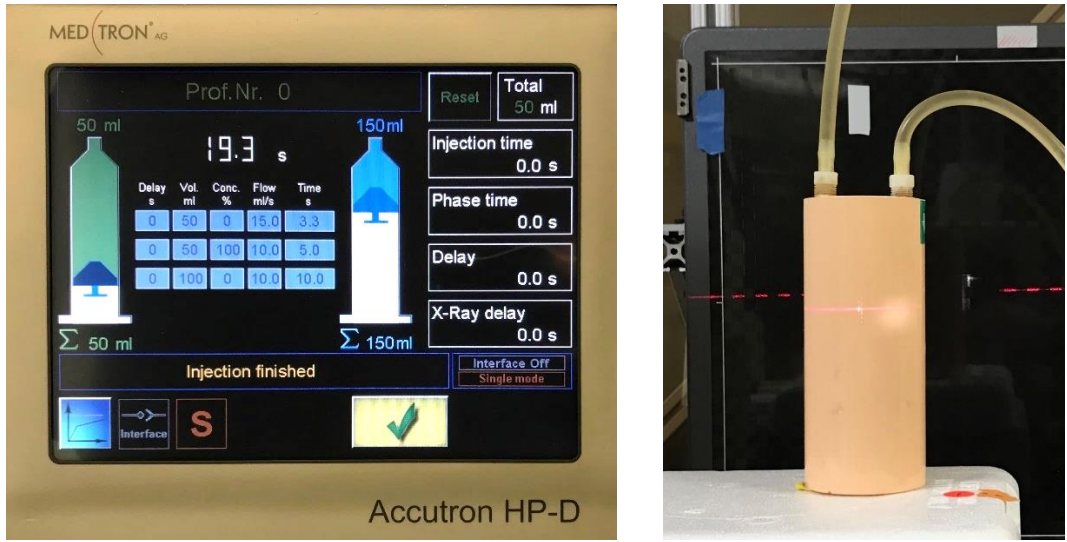

**Figure 7.** Experimental setup for dynamic flow phantom. Left: Power injector interface, showing contrast injection protocol for the flow phantom and post-injection results. Right: Dynamic flow phantom with U-shaped tube embedded in a solid water cylinder, shown against the background of the DL detector used to capture the dynamic image sequence.

## 3. Results

### 3.1 Simulation

Figure 8 shows the MD results using the proposed SSQI algorithm on the simulated data, and demonstrates the impacts of PM pitch and focal spot blur on the performance of SSQI. As compared to the ground-truth acrylic and copper components of the simulated data, without SSQI or scatter correction, the root mean squared error (RMSE) after MD was 11.35 mm for acrylic and 0.71 mm for copper, including the errors from residual bone in the acrylic image, soft tissue in the bone image, and residual PM pattern. SSQI removes most bias (minor errors pointed to by yellow arrows), with improved performance for smaller PM pitch and reduced focal spot blur. The soft tissue images clearly reveal the ground glass opacities caused by COVID-19 infection, as pointed to by green arrows. The RMSE was reduced to 5.72, 4.08, and 3.36 mm for acrylic and 0.18, 0.13, and 0.11 mm for Cu, respectively, clearly demonstrating the benefits of small focal blur (increased SMD) and finer sampling of PM (smaller PM pitch).

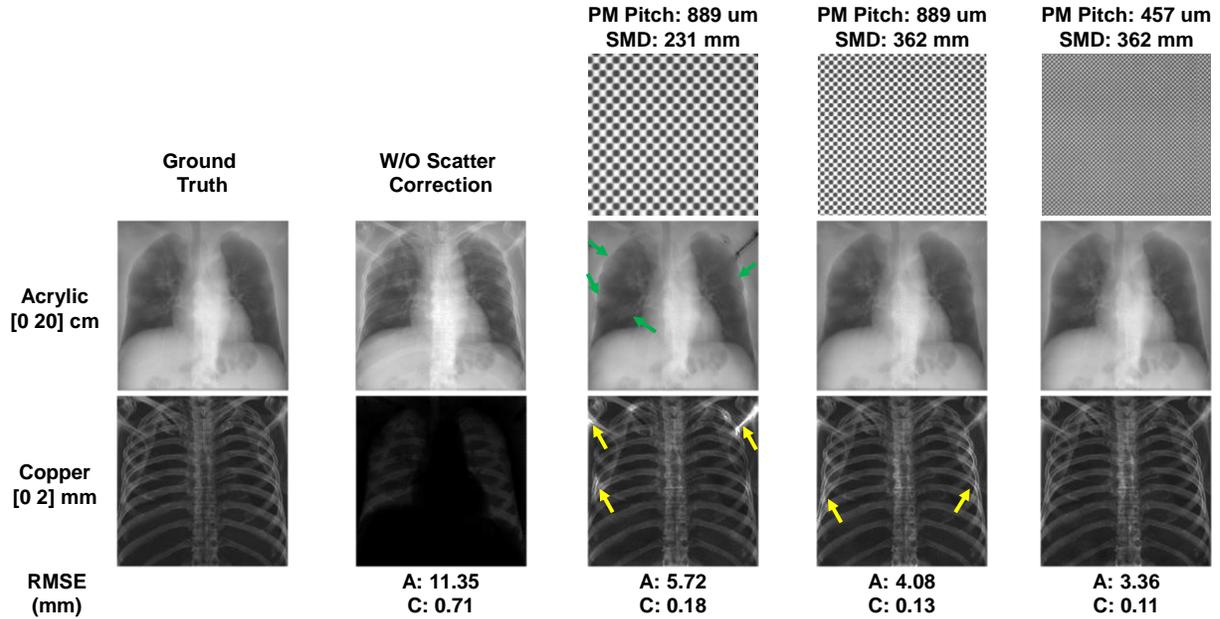

**Figure 8.** The impacts of PM pitch and focal spot blur on the performance of SSQI, which improved with smaller focal blur and smaller pitch size.

### 3.2 Calibration and Validation

To test the quantitative accuracy of SSQI, we used the validation dataset for testing and compared the derived MD with ground truth. The root mean squared error (RMSE) of SSQI estimation was 0.13 cm for acrylic and 0.04 mm for copper. The maximum absolute error was 0.23 cm for acrylic (1.02%) and 0.07 mm for copper (7.78%) for the thickest combination of 22.5 cm acrylic and 0.9 mm copper.

### 3.3 Static SSQI

With confidence from our validation study, we further tested the SSQI performance using the anthropomorphic chest phantom containing barium contrast. As shown in Figure 9, without applying SSQI or scatter correction, direct MD results in incorrect interpretation of acrylic and copper components. SSQI successfully separates soft-tissue and high-Z regions without leaving any residual modulator pattern. As compared to the ground truth MD within the collimated 2×2 cm$^2$ central region of the phantom, the RMSE was reduced from 0.45 cm to 0.28 cm for acrylic (38% reduction) and from 0.25 mm to 0.02 mm for copper (92% reduction). These results were achieved using a PM pitch of 457 µm and SMD of 36.2 cm, i.e., finer PM pitch and smaller focal blur, and the errors were consistent with our validation slabs.

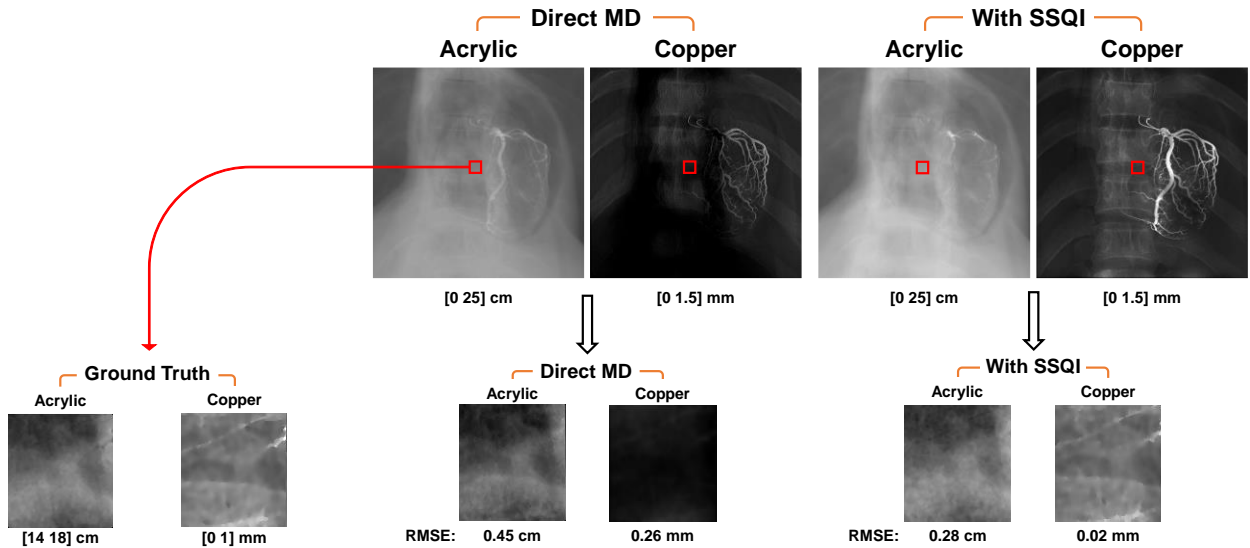

**Figure 9.** Comparison of MD with and without SSQI. Top row: With direct MD (left), large errors are seen in the MD maps as part of the bones and contrast are not revealed in the copper map. With SSQI (right), the MD is performed correctly. Bottom row: Zoomed-in comparison between ground truth and estimates in red regions.

### 3.4 Dynamic SSQI

Figure 10 shows the performance of SSQI for quantifying dynamic images of iodine contrast as a function of time. Without SSQI or scatter correction, the residual modulator patterns are clearly seen due to the scatter. SSQI successfully removes these patterns and clearly displays the iodine contrast distribution in each frame. Figure 10 also shows the comparison of flow curves measured at different locations within the flow tube as a function of time. The median values of the three square ROIs in different colors were used to plot the flow curves. SSQI provided quantitative measurements of contrast concentrations and captured the flow delay precisely. Since the injected iodine concentration was known to be 150 mg/ml through a 6.35 mm thick tube, the theoretical peak value for iodine contrast for an acrylic/copper MD was calculated to be 0.49 mm copper, which is slightly higher than but close to the results seen in the flow curves.

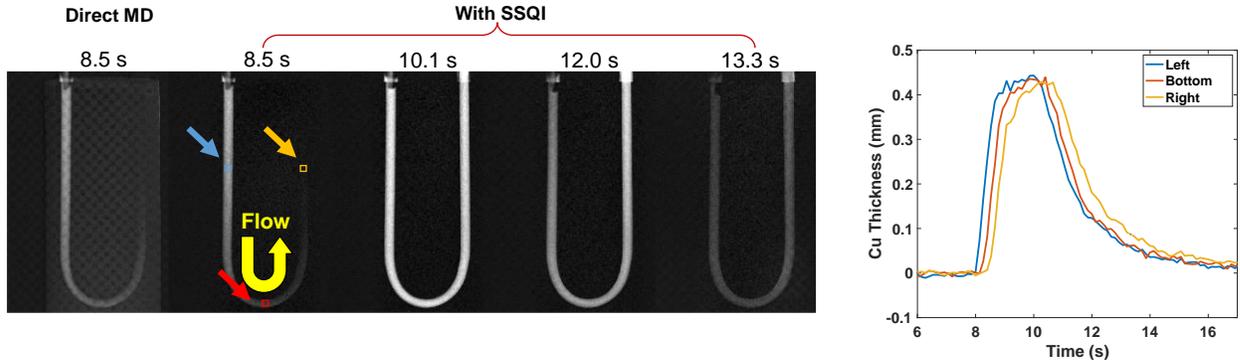

**Figure 10.** (Left) Copper images shown as a function of time. The far-left image was the result of direct MD without scatter correction, where the residual modulator pattern is clearly seen due to scatter. SSQI successfully removed these patterns and clearly displayed dynamic flow of the iodine contrast. Window: [-0.05 0.50] mm. (Right) Comparison of SSQI flow curves measured at different locations as a function of time. The median values of the three square ROIs (in three different colors that match the curves) were used to plot the flow curves. SSQI accurately provided quantitative measurements of contrast concentrations and captured the flow delay.

## 4. Discussion and Conclusion

In this work, we proposed a new SSQI method and algorithm that jointly estimates scatter and performs MD with high accuracy and efficiency. SSQI combines the use of a PM and DL detector, with each component serving its original function while their limitations are compensated by each other. We first demonstrated the feasibility using simulation studies performed on a COVID patient scan, and then further verified its efficacy with an anthropomorphic chest phantom and a dynamic flow phantom. Compared to direct material decomposition, SSQI greatly improved the MD accuracy. In addition to directly using the MD images, they can be combined to form a virtual monoenergetic (grayscale) image, as is typically done in DE imaging. While the goal of this work was to demonstrate initial feasibility, a number of parameters could be optimized in the future, including the PM pitch, thickness, and material, SMD, and DL detector design. MD calibration could be further improved by considering spatially-varying source spectra (e.g., heel effect) since we applied the central ray calibration to the entire image. In addition, MD calibration could be performed using more thickness pairs.

In ongoing work, we are also interested in investigating SSQI performance on clinical applications requiring quantitative assessment. For example, trans-arterial chemoembolization (TACE) is the most widely used therapy for hepatocellular carcinoma (HCC) [29,30]. The success of TACE critically depends on x-ray image guidance, and SSQI may provide quantitative guidance and visualization to determine the appropriate drug quantity to be delivered and delivery endpoints.

SSQI is potentially a new x-ray imaging scheme in high-resolution large-FOV imaging that brings quantitation to every view by simultaneously overcoming several challenges of x-ray imaging. It can be broadly applied to various x-ray systems, including radiography, fluoroscopy, angiography (e.g., maskless DSA), and CBCT systems. To the end user, SSQI can enable a seamless transition from conventional imaging, providing quantitative images without an additional imaging protocol.

**Acknowledgements:** This work was supported by NIH R21EB030080 and Varex Imaging.

**References**

1. Kelcz F, Zink FE, Peppler WW, Kruger DG, Ergun DL, Mistretta CA. Conventional chest radiography vs dual-energy computed radiography in the detection and characterization of pulmonary nodules. Am J Roentgenol 1994;162(2):271–8.
2. Brody WR, Cassel DM, Sommer FG, Lehmann LA, Macovski A, Alvarez RE, et al. Dual-energy projection radiography: Initial clinical experience. Am J Roentgenol 1981;137(2):201–5.
3. Siewerdsen JH, Moseley DJ, Bakhtiar B, Richard S, Jaffray DA. The influence of antiscatter grids on soft-tissue detectability in cone-beam computed tomography with flat-panel detectors. Med Phys 2004;31(12):3506–20.
4. Tang X, Ning R, Yu R, Conover DL. Investigation into the influence of x-ray scatter on the imaging performance of an x-ray flat-panel imager-based cone-beam volume CT. In: Medical Imaging 2001: Physics of Medical Imaging. 2001. page 851.
5. Sechopoulos I. X-ray scatter correction method for dedicated breast computed tomography. Med Phys 2012;39(5):2896–903.
6. Wiegert J, Bertram M, Rose G, Aach T. Model based scatter correction for cone-beam computed tomography. In: Flynn MJ, editor. Medical Imaging 2005: Physics of Medical

Imaging. 2005. page 271.

7. Siewerdsen JH, Daly MJ, Bakhtiar B, Moseley DJ, Richard S, Keller H, et al. A simple, direct method for x-ray scatter estimation and correction in digital radiography and cone-beam CT. Med Phys 2006;33(1):187–97.

8. Sun M, Star-Lack JM. Improved scatter correction using adaptive scatter kernel superposition. Phys Med Biol 2010;55(22):6695–720.

9. Wang A, Maslowski A, Messmer P, Lehmann M, Strzelecki A, Yu E, et al. Acuros CTS: A fast, linear Boltzmann transport equation solver for computed tomography scatter – Part II: System modeling, scatter correction, and optimization. Med Phys 2018;45(5):1914–25.

10. Zbijewski W, Beekman FJ. Efficient Monte Carlo based scatter artifact reduction in cone-beam micro-CT. IEEE Trans Med Imaging 2006;25(7):817–27.

11. Colijn AP, Beekman FJ. Accelerated simulation of cone beam X-Ray scatter projections. IEEE Trans Med Imaging 2004;23(5):584–90.

12. Shi L, Vedantham S, Karellas A, Zhu L. Library-based scatter correction for dedicated cone beam breast CT: a feasibility study. In: Kontos D, Flohr TG, Lo JY, editors. Medical Imaging 2016: Physics of Medical Imaging. 2016. page 978330.

13. Maier J, Sawall S, Knaup M, Kachelrieß M. Deep Scatter Estimation (DSE): Accurate Real-Time Scatter Estimation for X-Ray CT Using a Deep Convolutional Neural Network. J Nondestruct Eval 2018;37(3).

14. Zhu L, Bennett NR, Fahrig R. Scatter correction method for X-ray CT using primary modulation: theory and preliminary results. IEEE Trans Med Imaging 2006;25(12):1573–87.

15. McCollough CH, Leng S, Yu L, Fletcher JG. Dual- and multi-energy CT: Principles, technical approaches, and clinical applications. Radiology 2015;276(3):637–53.

16. Patel R, Panfil J, Campana M, Block AM, Harkenrider MM, Surucu M, et al. Markerless motion tracking of lung tumors using dual-energy fluoroscopy. Med Phys 2015;42(1):254–62.

17. Shi L, Bennett NR, Wang AS. Single-shot quantitative x-ray imaging using a primary modulator and dual-layer detector: simulation and phantom studies. In: SPIE Medical Imaging. 2022. page 24.

18. Shi L, Bennett NR, Shiroma A, Sun M, Zhang J, Colbeth R, et al. Single-pass metal

artifact reduction using a dual-layer flat panel detector. Med Phys 2021;48(10):6482–96.
19. Müller K, Datta S, Ahmad M, Choi JH, Moore T, Pung L, et al. Interventional dual-energy imaging - Feasibility of rapid kV-switching on a C-arm CT system. Med Phys 2016;43(10):5537–46.
20. Haytmyradov M, Mostafavi H, Wang A, Zhu L, Surucu M, Patel R, et al. Markerless tumor tracking using fast-kV switching dual-energy fluoroscopy on a benchtop system. Med Phys 2019;46(7):3235–44.
21. Tivnan M, Wang W, Stayman JW. A prototype spatial–spectral CT system for material decomposition with energy-integrating detectors. Med Phys 2021;48(10):6401–11.
22. Shi L, Lu M, Bennett NR, Shapiro E, Zhang J, Colbeth R, et al. Characterization and potential applications of a dual-layer flat-panel detector. Med Phys 2020;47(8):3332–43.
23. Wang AS. Single-shot quantitative x-ray imaging from simultaneous scatter and dual energy measurements: a simulation study. In: SPIE Medical Imaging. 2021. page 77.
24. Gao H, Zhu L, Fahrig R. Modulator design for x-ray scatter correction using primary modulation: Material selection. Med Phys 2010;37(8):4029–37.
25. Coleman TF, Li Y. An interior trust region approach for nonlinear minimization subject to bounds. SIAM J Optim 1996;6(2):418–45.
26. Shi L, Vedantham S, Karellas A, Zhu L. X-ray scatter correction for dedicated cone beam breast CT using a forwardprojection model. Med Phys 2017;44(6):2312–9.
27. Cohen JP, Morrison P, Dao L. COVID-19 Image Data Collection. arXiv:200311597
28. Meyer E, Maas C, Baer M, Raupach R, Schmidt B, Kachelries M. Empirical scatter correction (ESC): A new CT scatter correction method and its application to metal artifact reduction. In: IEEE Nuclear Science Symposium Conference Record. 2010. page 2036–41.
29. Bertuccio P, Turati F, Carioli G, Rodriguez T, La Vecchia C, Malvezzi M, et al. Global trends and predictions in hepatocellular carcinoma mortality. J Hepatol 2017;67(2):302–9.
30. Llovet JM, Ducreux M, Lencioni R, Di Bisceglie AM, Galle PR, Dufour JF, et al. EASL-EORTC Clinical Practice Guidelines: Management of hepatocellular carcinoma. J Hepatol 2012;56(4):908–43.